# Think twice - it's worth it!

# Improving the performance of minority games

J. Menche[1,2] and J.R.L de Almeida[1]

[1]Departamento de Física, Universidade Federal de Pernambuco, 50670-901, Recife, PE, Brasil
almeida@df.ufpe.br
[2]Fakultät für Physik und Geowissenschaften, Universität Leipzig, Linnéstraße 5, 04103 Leipzig, Germany
joergmenche@gmx.de

ABSTRACT

In this work the properties of multi choice minority games (MCMG) are studied by means of extensive computational simulations. We have considered several ways of rewarding the player's strategies and compared the resulting behaviours of the ensemble. It is shown that introducing a reward for the second place can improve the overall performance significantly. Our new but still very simple architecture of strategies shows interesting properties. In the version of the MCMG which also rewards second best choice the unordered phase is strongly suppressed and the global performance of the ensemble is significantly improved. The efficiency of this model can come close to the one of standard minority games (SMG) that use adaptive genetic algorithms for searching good strategies.

## I. Introduction

The minority game (MG) is a very simple model of interacting agents. Nevertheless it shows fascinating properties and poses novel challenging questions to Statistical Mechanics and its methods and ideas. In short, its main idea is the following: an odd number of selfish players repeatedly has to choose one out of two possible actions. These could be for example buying or selling at the stock market, taking the bus or the metro, which internet router is the fastest (ie. less crowded). As it will become clearer in the examples, the players that are in the minority side will win the game. If there are more sellers than buyers, the price will fall and the buyers will be in advantage. If there are lots of people in the metro it would have been much more comfortable to take the bus and so on. This kind of game was inspired by Arthur's ``El Farol'' problem [1]. There it was an overcrowded bar that rose his interest of how a population of heterogeneous players interacts.
In this section we give a short and necessarily incomplete summary of the SMG and the MCMG and their well-known properties. Then we shall present our models of the MCMG and compare them with the standard models.
In the SMG that was introduced in [2,3], $N$ agents (or players) choose at each time step $\tau$ one out of two possible options, say 0 and 1. Agents that belong to the smaller of both groups win. They act completely independent, without any possibility of communication or interaction. Their only available information about the other player's actions is a ``history" (or ``memory") of the last $m$ games' winners. There are two possible values for each of the $m$ entries of the history, so there are $2^m$ possible different histories. Every player is equipped with a set of $S$ strategies to determine their next choice. As the decision depends on the current history of the last games, these strategies contain $2^m$ values, each of them representing the choice for one of the possible histories. All of the $2^m$ entries can be 0 or 1, so there are $2^{2^m}$ different strategies. After all agents having made their choice and the winning group being determined, the agents compare the predictions that their strategies made with the present outcome. If a strategy was right, it will be rewarded with one point. The agents will always use the best strategy, that means the one with the most points, to take their choice. An interesting quantity in the MG is the standard deviation

$$\sigma = \sqrt{\sum_{i=0}^{R} \frac{1}{R}\left(N_i^{(1)} - \frac{N}{2}\right)^2}$$

where $R$ is the total number of rounds, $N_i^{(1)}$ is the number of agents playing 1 (or 0) in round $i$. The standard deviation is a measure for the efficiency of the system at distributing the limited resources. $\frac{N}{2}\left(\approx \frac{N-1}{2}\right)$ is the maximum number of points to be won per time step. Small variations around this value mean therefore, that the system is very efficient. Fig.1 shows our results for two different values of $N$.
Numerical simulations as fig.1 or in [4], as well as analytical calculations [5] show that the system undergoes a phase transition. The two phases are separated by the minimum of the curve. On the left side of the minimum we observe that the value of $\sigma$ varies widely from the medium value, whereas on the right side these variations vanish almost completely. In [6] it is shown that all the curves for different values of $N$ and m fall on a universal curve by using a $\sigma^2/N$ vs. $2^m/N$ plot, see fig.2. For large values of the ratio $\alpha = 2^m/N$ the system displays a cooperative phase ($\alpha > \alpha_c \approx 0.34$), for small values ($\alpha < \alpha_c$) the coordination of the agents is worse than the one of a ``random system``. The ``random system" consists of agents that make their choice by coin toss, without using any strategy.

These general properties do not depend on the number of available strategies $S$ that the agents can use. This is shown in fig.3 for S=4.
In fig.4 we directly compare the performances of two ensembles with S=2 and S=4. There we see that four strategies are less efficient than two strategies, as it is explained in [6]. Only for larger values of $m$ (almost random choice) the results are the same.
The standard minority game with two choices can be generalized to more than just two options [7].
In fact life would be kind of sad, if there always were only two options to choose from. The qualitative behaviour remains the same as it is for two choices. Only the value $\alpha_c$ where the phase transition occurs depends on the number of choices. This is valid in case that only a first place is rewarded, ie., only the smallest minority is rewarded.

In the following section we'll present our models to reward also a second place. We thought this is just fair and wanted to find out, whether this fairness shows any effect on the behaviour of the ensemble.

## II. Models

So getting the silver medal isn't that bad. It is better than not getting anything at all. We wanted to introduce this idea in the minority game and analysed three different ways of doing so:

(1) Every agent is equipped with S normal strategies like in the SMG. If a strategy led to a first place it will be rewarded with *f* points. If the strategy didn't led to a first place, but at least to a second place, it will be rewarded with *s* points. Therefore the maximum amount of points a strategy can win per round is *f*.

(2) Every agent is equipped with S strategies that make two predictions at one time, one for the first place and one for the second place. The agents will always use the prediction for the first place to make their choice. If a strategy made a right prediction for the first place, it will be rewarded with *f* points. If it made a right prediction for the second place, it will be rewarded with s points. Because the strategies make two predictions they can gain *f*+*s* points here.
We introduce a separate history for the second places of the *m* last rounds. The strategies make their prediction for the second place based on this history.

(3) The same as (2), but in this case both predictions are based on the history of the first place.

Let us give a short summary of the parameters that occur in our models: Number of agents *N*, length of history *m*, number of strategies *s*, reward for a first place *f*, reward for a second place *s*.

## III. Results

### A. Comparison between the standard deviations

First we studied the behaviour of the standard deviation. In fig.5 and 6 we show the results for different values of *f* and *s*.
For all of our models we carried out simulations with *f=3, s=2* and with *f=5, s=0* to check the influence of rewarding the second place. In the uncoordinated phase ( $m \geq 4$ for S=2; $m \geq 5$ for S=4) there's no difference, all the curves fall on one. For smaller values of *m*, however, there's a big difference. The results of model (1) are invariant under introducing *s*. The models (2) and (3) on the contrary show a significantly improved performance for $s \neq 0$. In these cases the uncorrelated phase of the MG is strongly suppressed and the system remains in the correlated phase. We find this very interesting, because it means that the overall performance of these systems is much better. They show only small fluctuations around the optimal value of $N_{max}$. In the sense of the global win of all agents they are therefore more efficient, because more agents win at the same time.

**B. Scaled curves**

To study the behaviour of our models and its phases, we plotted $\sigma^2/N$ vs. $3^m/N$ (following [8]) in fig.7, 8 and 9. In the models (2) and (3) that reward the second place the uncorrelated phase is so strongly suppressed that we needed to go up to very high values of N, until $\sigma^2/N$ reached and than finally climbed over the value of pure random choice. For model (1) this value lies at $\alpha \approx 0.095$, for model (2) at $\alpha \approx 0.0012$ and for model (3) at $\alpha \approx 0.01$. For $m$ fixed at m=2 these values correspond to $N \approx 95, N \approx 7500$ and $N \approx 900$, which is a considerable difference.

To confirm this we directly compare the performances of the models (2) and (3) with the one of the standard model in fig.10. There it is shown clearly that introducing a reward for the second place improves the performance significantly. The minimum value of $\sigma^2/N$ at $\alpha_c$ is smaller and the phase transition occurs earlier.

**C. The scaled utility**

To get a better insight into the efficiency of our models we study the scaled utility function $u$ that was introduced in [9]: $u(N_i)=N_i/N_{max}$, where $N_i$ is the number of winners in round $i$, and $N_{max} \approx \frac{N}{Q}$ ($Q$ is the number of options to choose from) is the maximum number of agents that can win at the same time. $u \cong 1$ means that the distribution of the limited resources is optimal, $u<1$ means, that only few agents won. Their personal gain might be higher, but the system in the whole is less efficient. In section **III A** we showed that the model (2) with $f=3$, $s=2$ has a much better performance than the models that don't reward second places. In fig.11 we compare u of model (1) ($f=5$, $s=0$), model (2) ($f=3$, $s=2$) and of the SMG with two choices for $m=2$, $m=5$, $m=8$.

For $m=2$ the performance of the MCMG is significantly better than the one of the SMG. The fluctuations are much smaller and the value of $u$ is nearer to 1, meaning that the resources of the system are better distributed. For $m=5$ the scaled utility functions of the SMG and the MCMG are almost the same, only the fluctuations of the MCMG are smaller. For $m=8$ the SMG has a better performance.

To make this clearer we calculated for each value of m the mean of the scaled utility function $u$ for all the different models and put them together in

For $m=2$ and $m=3$ the models(1) and (2) that include rewarding for the second place, are by far better, than models that just reward the first place, as well as the models that offer only two choices. This corresponds with the results in section **III A** and **III B**. At $m=4$ all the MCMG graphs unite, but are still above the graph of the two choice MG. This changes for $m \geq 6$, from there on the standard minority game has a better performance.
This is reasonable, because in this range the systems are already near the pure random choice. But with the increase of choices increases the possibility to fail when tossing a coin (or rather throwing a dice for more than two options).

## IV. Conclusions

By introducing a history of the second places we give more information to the agents. In the SMG the only public information are the recent first places, so in our models the agents know more about the state of the game and about the quality of their strategies. In contrary to proceeding to the next higher m, where the strategy space increases exponentially it only increases by factor two in our case. Nevertheless the agents can explore this additional information considerably and the performance of the system becomes significantly better. The best performance is shown by model (2), where we put the exact history of the second places as additional information into the game. The phase transition into the uncorrelated phase is strongly suppressed. In model (3) the strategies use the history of the first places to predict the next outcome for the first and the second place, so they don't have exact information about the development of the second places. Therefore the performance is poorer as for model (2), but still better than without considering the second best choice. In Model (1) no such improvement is shown, as there's no additional information at all that the agents could explore.
Apart from analysing the standard deviations we studied the properties of the scaled utility function to confirm our conclusions. We found that the performance of models that take the second place into account can come close to the one of SMG that uses genetic algorithms to find optimal strategies.
It seems worthwhile to try to find analytical models to describe and explain our numerical results. In [5] it is shown, that the SMG can be described as a spin system. It should be possible to generalize these results to more than just two options. The generalization from SMG to MCMG can be seen as the generalization from the Ising model to the Potts model. But unfortunately, as mentioned in [8], this analogy doesn't help directly in calculating the properties of the MCMG. As far as we know there's no solution found up to now. Speaking in the spin formalism, introducing a reward for the second place could be interpreted as an external field, that disturbs the symmetry of the system and seems by our numerical results to simulate an ordering field for this complex system
The minority game currently rises much interest in the physical community and there are a lot of suggestions and ideas, what could be done next. We find it especially fascinating to study effects that can be regarded to have an analogy in the real world out there. What happens to the whole system, if we reward the second place? As we could show in this paper, the performance of the system is significantly improved, in our case about 50 %.

## Acknowledgments

J.Menche wishes to thank the DAAD (German academic exchange service).

---

[1] W. B. Arthur, Amer. Econ. Review 84, 406 (1994).
[2] Y.-C. Zhang, Europhys. News 29, 51 (1998).
[3] D. Challet and Y.-C. Zhang, Physica A 246, 407 (1997).
[4] R. Savit, R. Manuca and R. Zecchina, Phys. Rev. Let. 82, 2203 (1998).
[5] D. Challet, M. Marsili, and R. Zecchina, Phys. Rev. Let.84, 1824 (2000).
[6] D. Challet and M. Marsili, Phys. Rev E 60, R6271 (1999).
[7] L. Ein-Dor, R. Metzler, I. Kanter, and W. Kinzel, Phys Rev. E63, 066103 (2001).
[8] R. Metzler, Ph.D. thesis, Universität Würzburg (2002).
[9] M. Sysi-Aho, A. Chakraborti, and K.Kaski (2003), cond-mat/0305283v1.

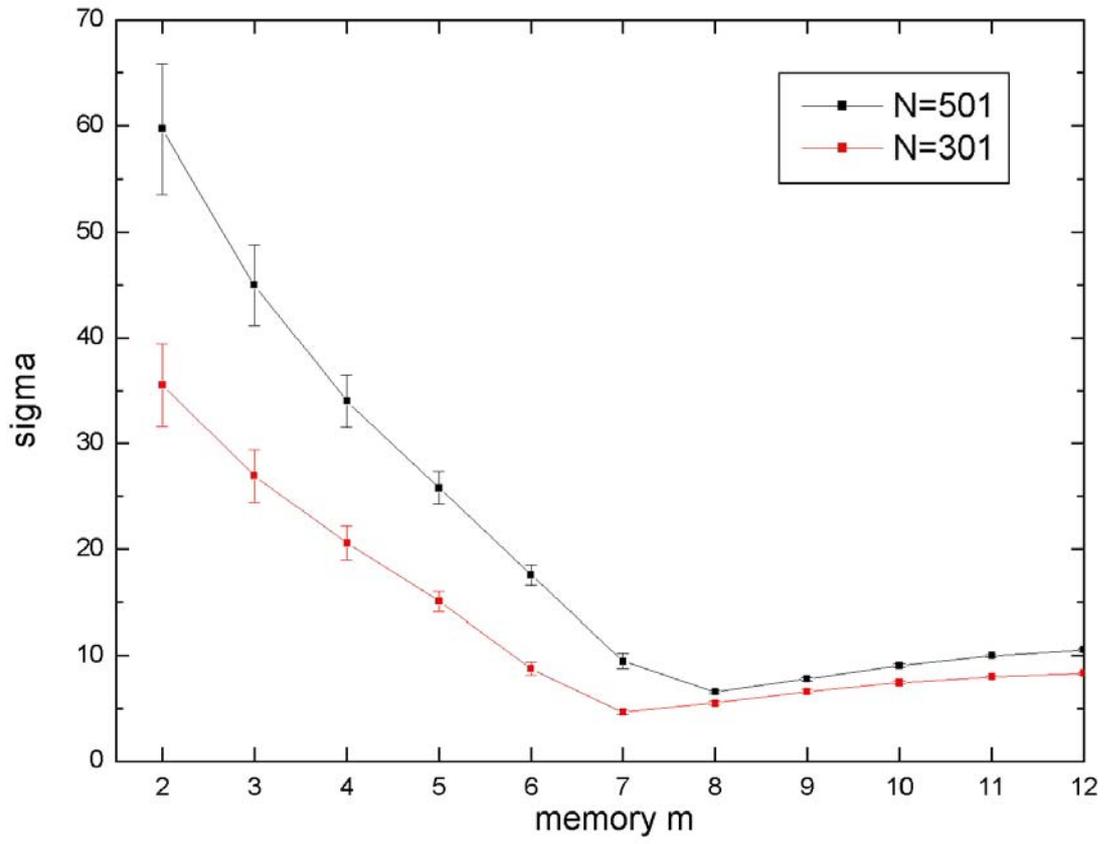

FIG. 1: SMG; $\sigma$ vs m ; S = 2

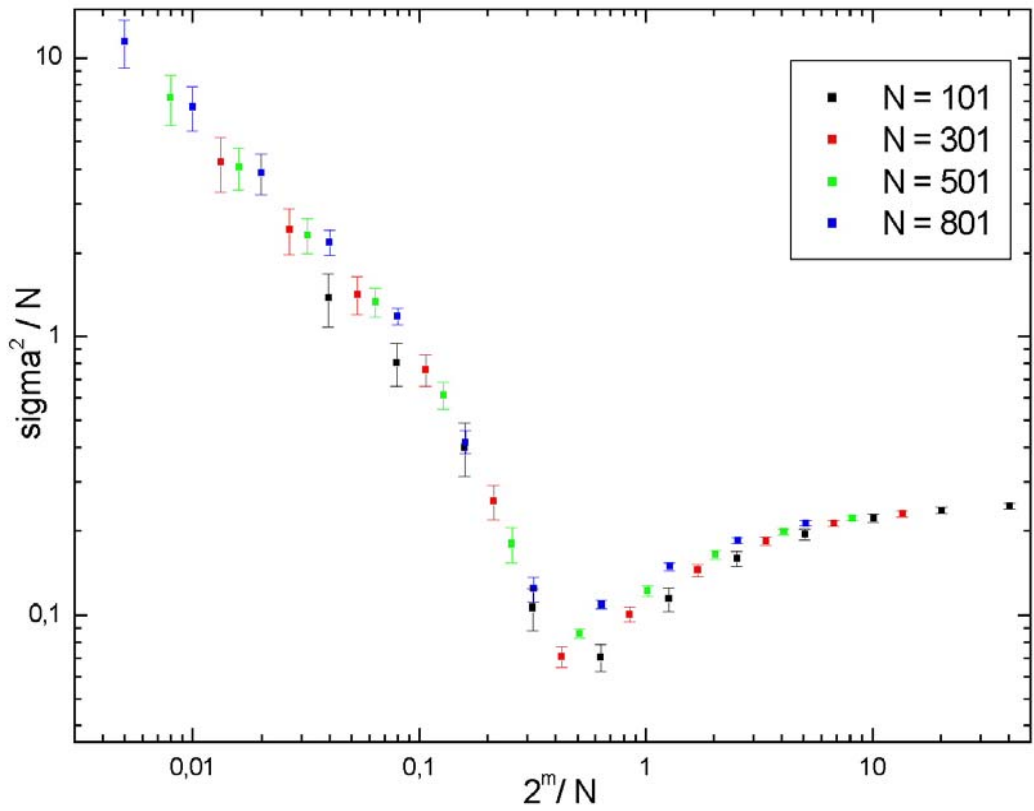

FIG. 2: SMG; $\sigma^2 / N$ vs $2^m / N$ ; S = 2

FIG. 3: SMG; $\sigma^2 / N$ vs $2^m / N$ ; S = 4

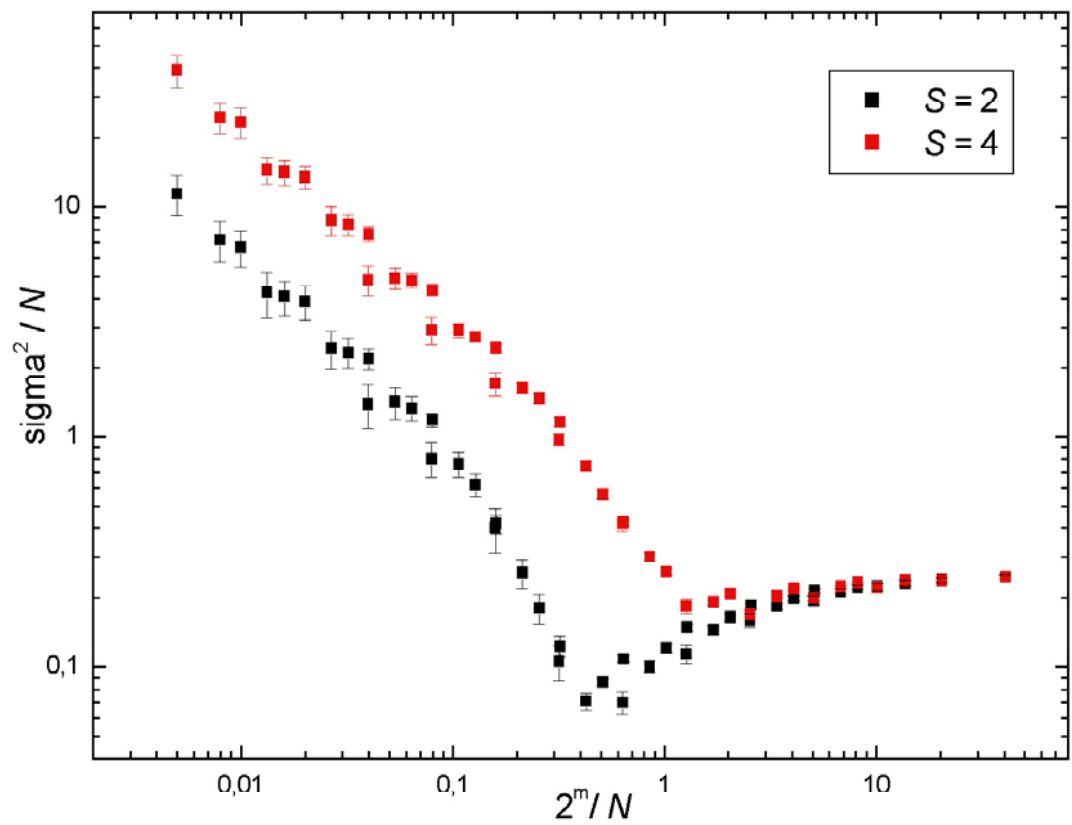

FIG. 4: SMG; $\sigma^2 / N$ vs $2^m / N$ ; S = 2 & S = 4

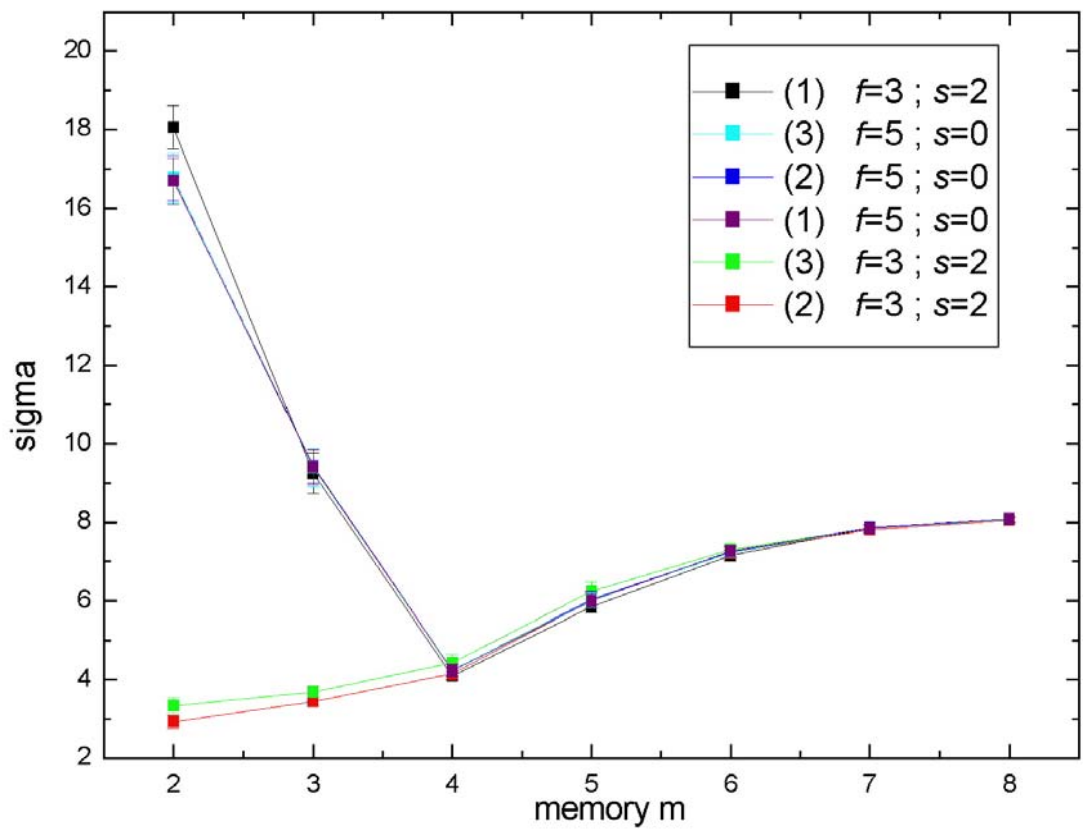

FIG. 5: MCMG; σ vs m ; S = 2; N = 301

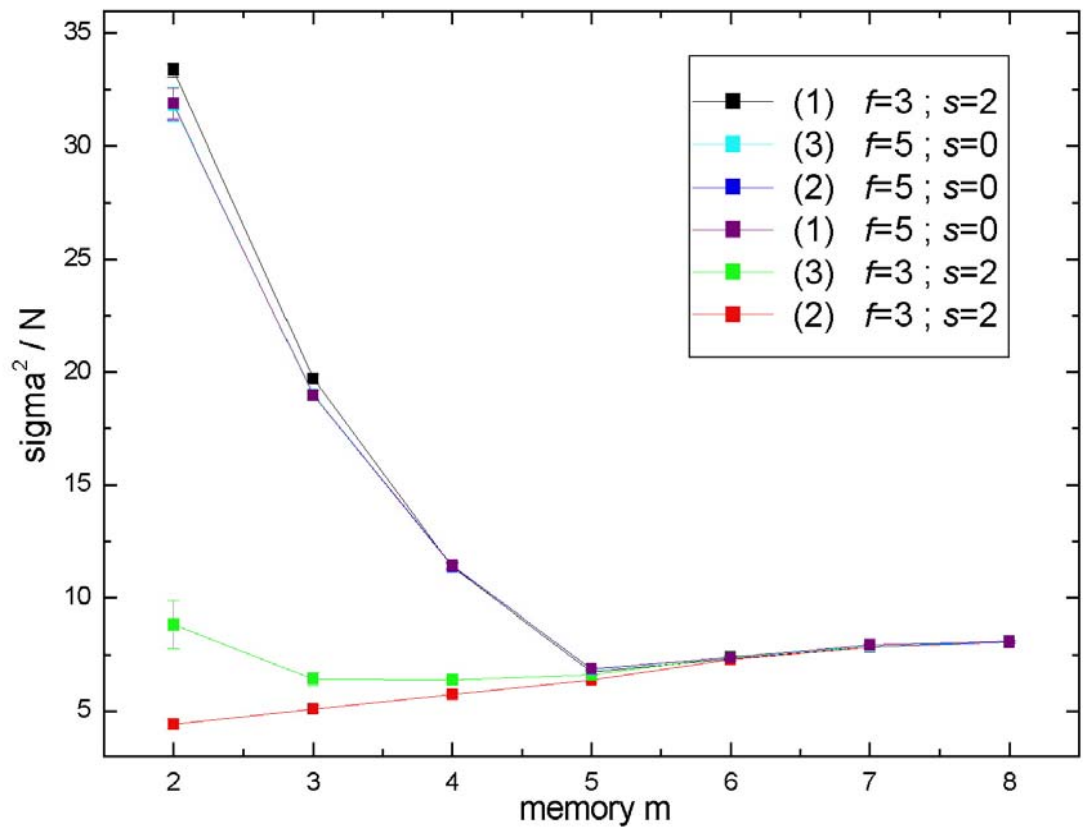

FIG. 6: MCMG; σ vs m ; S = 4; N = 301

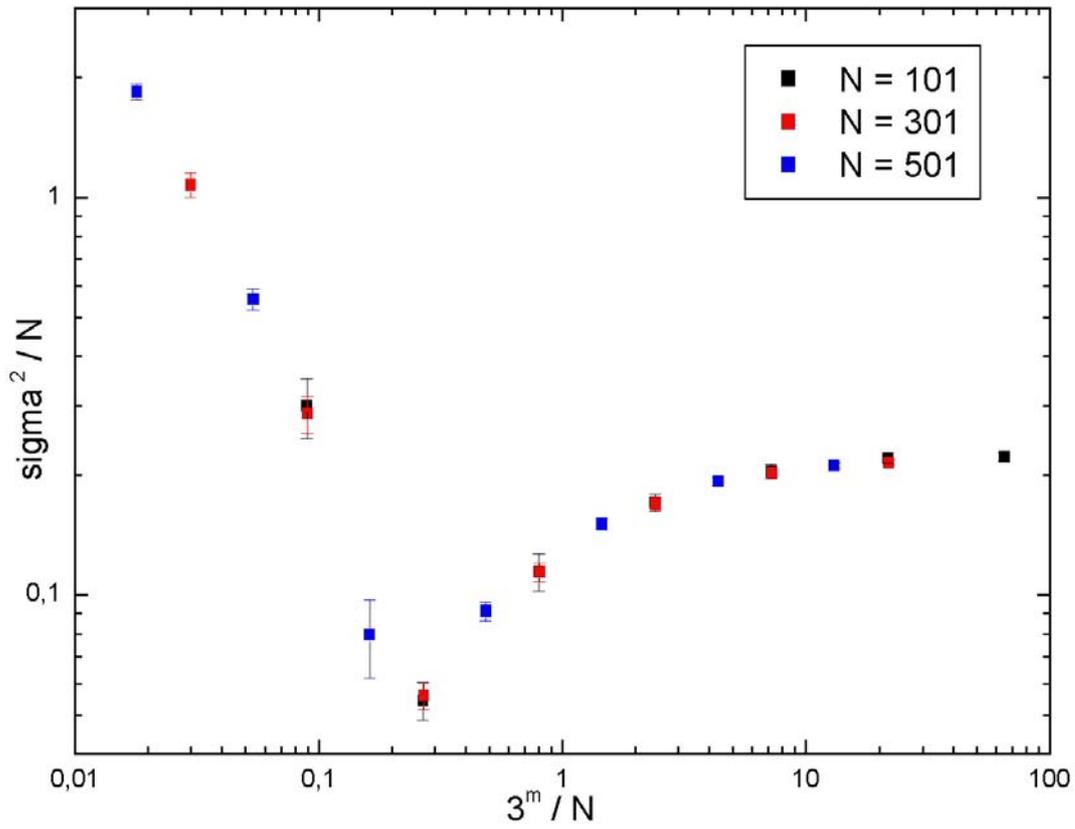

FIG. 7: Model (1) : $\sigma^2 / N$ vs $3^m / N$ ; f = 3; s=2; S = 2

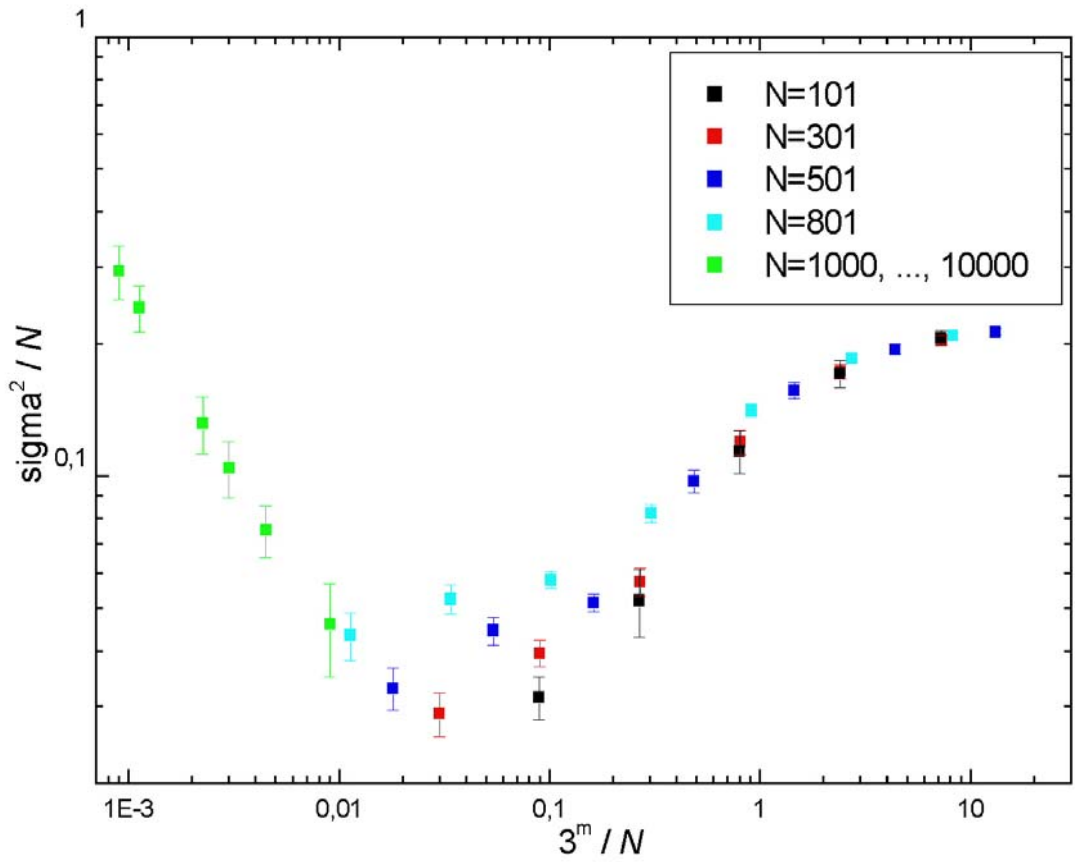

FIG. 8: Model (2) : $\sigma^2 / N$ vs $3^m / N$ ; f = 3; s=2; S = 2

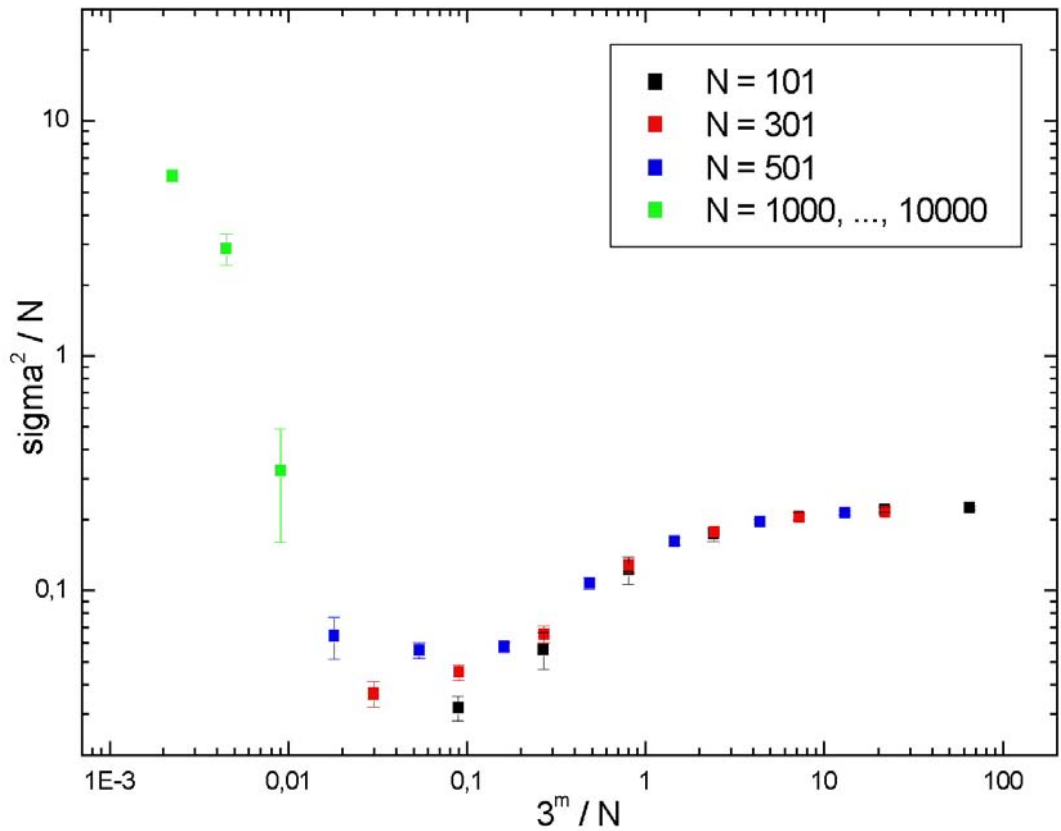

FIG. 9: Model (3) : $\sigma^2 / N$ vs $3^m / N$ ; $f = 3$; $s=2$; $S = 2$

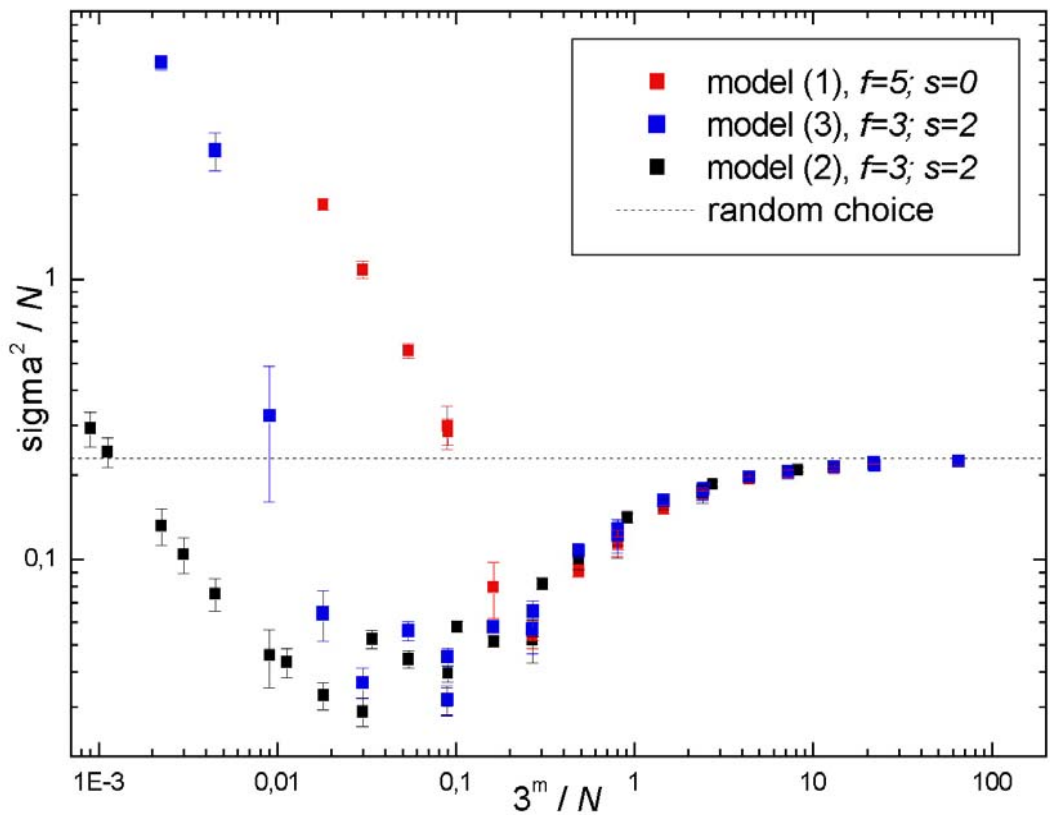

FIG. 10: $\sigma^2 / N$ vs $3^m / N$  for models (1),(2) and (3)

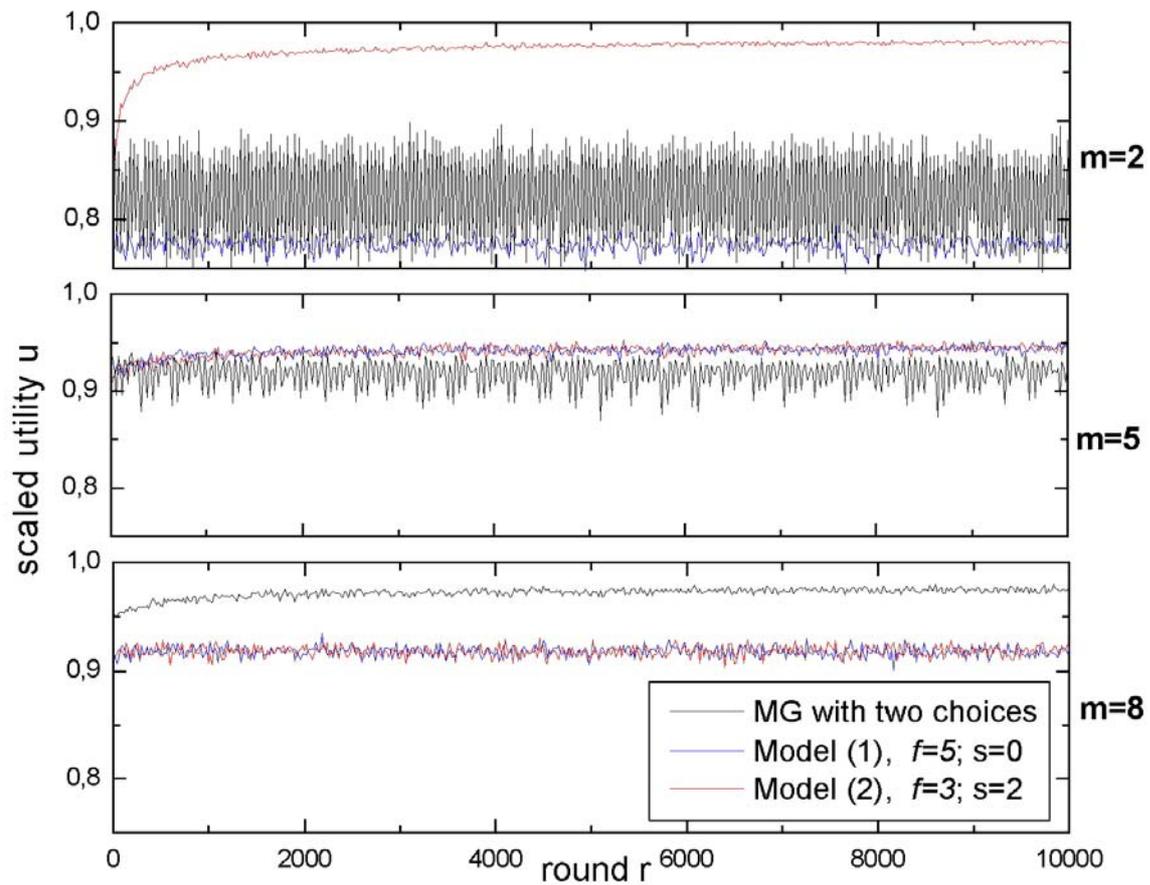

FIG. 11: $u$ vs $\tau$; $N = 301$; curves are average over 100 runs

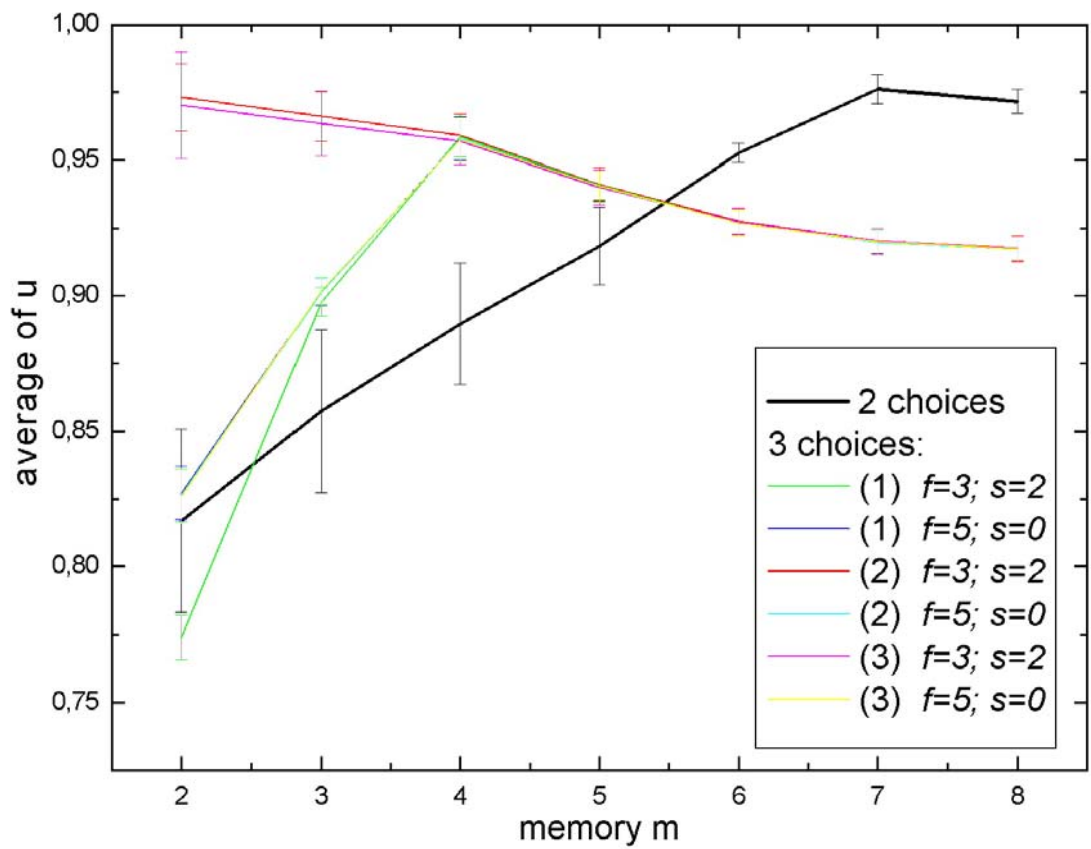

FIG. 12: $u$ vs $m$